\documentclass[12pt]{iopart}

\usepackage{iopams}  
\expandafter\let\csname equation*\endcsname\relax
\expandafter\let\csname endequation*\endcsname\relax
\usepackage{amsmath}
\usepackage{cite}
\usepackage{hyperref}
\hypersetup{
    colorlinks=true,
    linkcolor=blue,
    filecolor=blue,      
    urlcolor=blue,
    citecolor=blue
}
\usepackage{graphicx}

\usepackage{subcaption}
\DeclareMathOperator{\sinc}{sinc}

\begin{document}

\title[ ]{Experimental generation and characterization of partially spatially coherent qubits}

\author{Preeti Sharma\textsuperscript{$\dagger$}, Sakshi Rao\textsuperscript{$\dagger$} and Bhaskar Kanseri\textsuperscript{*}}

\address{Experimental Quantum Interferometry and Polarization (EQUIP), Department of Physics, Indian Institute of Technology Delhi, Hauz Khas, New Delhi 110016, India}
\ead{\textsuperscript{*}bkanseri@physics.iitd.ac.in}
\vspace{10pt}
\begin{indented}
\item[]Dec 2022
\end{indented}
\def\thefootnote{$\dagger$}\footnotetext{These authors contributed equally to this work}
\begin{abstract}
Partially spatially coherent qubits are more immune to turbulent atmospheric conditions than
coherent qubits, which makes them excellent candidates for free-space quantum communication. In this article, we report the generation of partially spatially coherent qubits in a spontaneous parametric down-conversion (SPDC) process using a Gaussian Schell model (GSM) pump beam. For this non-linear process, we demonstrate experimentally for the first time, the transfer of spatial coherence features of the pump (classical) to the biphotons (quantum) field. Also, the spatial profiles of partially coherent qubits generated in type-I and type-II non-collinear SPDC process are experimentally observed and multi-mode nature of partially coherent photons (qubit) is ascertained. These investigations pave the way toward the efficient generation of partially spatially coherent qubits with a tunable degree of spatial coherence, which lead to wide range of applications in frontier areas such as quantum cryptography, teleportation, imaging, and lithography.
\end{abstract}
\section{Introduction}
For decades, researchers have been intrigued by the prospect of free-space quantum communication, which offers secure information exchange using principles of quantum physics  \cite{peng2005experimental,aspelmeyer2003long,jin2010experimental}. However there are many challenges involved in free space communication, for instance, random variation of the refractive index of the atmosphere, which usually causes intensity fluctuations and beam wandering  
\cite{Churnsid90}. These factors degrade the quality of received information and limit the transmission distance to a great extent through strong atmospheric turbulence. 
An alternative method is to use partially coherent beams in such conditions, as they are robust against the deleterious effects of atmospheric turbulence \cite{dogariu2003propagation,gbur2014partially}. Additionally, the speckle pattern in these beams is diminished as a result of the reduced spatial coherence \cite{mandel1995optical,lajunen2011propagation}. Quantum light, such as single photons and entangled photons, can possess partial spatial coherence properties and may offer robustness against such turbulent conditions \cite{Shekel_2022,jha2010spatial}. Partially coherent beams have gained attention in the quantum realm due to their numerous applications in quantum communication, quantum metrology, and ghost imaging. 

There are many ways for the generation of the quantum state of light based on the second-order and third-order non-linearity of the medium \cite{boyd2020nonlinear}. The spontaneous parametric down-conversion (SPDC) process has been a well established method to generate entangled photons in the spatial, spectral, polarization, and hybrid domains. It also provides a means to generate true single photons by heralding one of the twin photons exhibiting quantum properties. It is a second-order nonlinear process where a photon of higher frequency is spontaneously down-converted into two photons (twin photons/biphoton) of lower frequency, named signal and idler. This process is governed by the energy and momentum conservation, or phase matching conditions \cite{gerry2005introductory}. 
These twin photons have been utilized in the investigation of fundamental features such as quantum interference, quantum cryptography, non-locality, and quantum teleportation. The study of correlations in spatial, spectral, and polarization domains with coherent pump is well explored in the literature. Conventionally, the process is categorized as type-0, type-I and type-II phase matching, based on the polarization of the interacting photons (pump, signal and idler). The polarization of signal and idler photons are identical in type-I and type-0 phase matching, with the difference that it is orthogonal to the pump in type-I case. On the other hand, in type-II phase matching, the signal and idler photons have orthogonal polarization. The transfer of the angular spectra of the pump to the biphotons has been experimentally observed with a collimated and focused coherent pump in type-I and type-II SPDC \cite{lee_exter_woerdman2005, Ram_rez_Alarc_n_2013}.  The transverse distribution of biphotons exhibits strong spatial correlations with the pump as a coherent beam. Lately, there have been some studies dealing with biphotons generated with a partially coherent and incoherent pump which revealed the dependence of pump coherence on profiles (theoretically) \cite{preeti2020}, quadrature squeezing \cite{sakshi2022}, quantum interference \cite{joshi2020spatial}, spatial entanglement \cite{defienne2019spatially, Zhang2019} and polarization entanglement \cite{preeti2021, zhang2022}. The study of interference in double-slit experiments using quantum beams has always been a topic of interest owing to quantum lithography \cite{Shimizu2003}, ghost imaging \cite{strekalov1995observation,li2019}, and fundamental studies \cite{Gatti2003,zhang2017}. The interference pattern in transverse positions of biphotons has been reported to tailor polarization entanglement \cite{Santos2001}. The coincidence detection in twin photons with double-slits placed in one \cite{Strekalov1995} or both paths \cite{Milena2001} have been reported with the finite size of coherent pump \cite{Ribeiro1997}. The transfer of coherence features from the pump to biphotons is of utmost relevance. It is demanding to recognize whether the quantum field generated in SPDC process and to be used in free-space quantum communication follows the similar statistical features as the well-documented classical pump beam. To the best of our knowledge, there is no experimental study so far revealing the transfer of the spatial coherence properties  of the pump (classical field) to biphotons (quantum field) in the SPDC process. It is thus quite important and timely to address this problem both experimentally and theoretically, resulting in the generation of tunable partially spatially coherent qubits.  

This work aims to investigate the coherence properties of photons generated using the SPDC process in a type-II non-collinear configuration. We experimentally explore the propagation of coherence features from the classical beam (pump) to the quantum beam (biphotons) in second-order non-linear interaction. The coherence properties of the pump are initially determined followed by the investigation of the coherence of photons generated using SPDC. The results show that the spatial coherence is transferred from the pump to biphotons and follows the same characteristics as the classical partially spatially coherent beam. These experimental findings are supported by the theoretical formulation provided in this paper. The study reveals that the partially coherent qubits with a tunable degree of spatial coherence to be used in free space optical communication can be easily generated using the proposed experimental configuration. The effect of changing the spatial coherence of the pump with a fixed beam size on the conditional detection probability and spatial profiles of biphotons generated in type-I and type-II non-collinear SPDC processes is also studied. The spatial profiles of the output beam show asymmetry with degradation in the spatial coherence of the pump. This would also have an impact on the use of these qubits for quantum lithography and imaging purposes.

This paper is organized as follows: In Section 2, we demonstrate the generation of a partially coherent GSM beam with a tunable degree of spatial coherence experimentally. 
In Section 3, we theoretically and experimentally investigate the effect of variation in the spatial coherence of the pump on the Young's double-slit interference in biphotons generated using the type-II SPDC process. In Section 4, the spatial profiles of photons in type-I and II non-collinear geometry are experimentally determined. 
The conditional probability distributions showing coincidence counts in type-II non-collinear SPDC for the entangled positions are determined with variations in pump coherence. In Section 5, the study is concluded.

\section{\label{sec:Theory-1}Generation and characterization of partially spatially coherent pump}
\begin{figure}[b!]
    \centering
    \includegraphics[width=0.9\columnwidth]{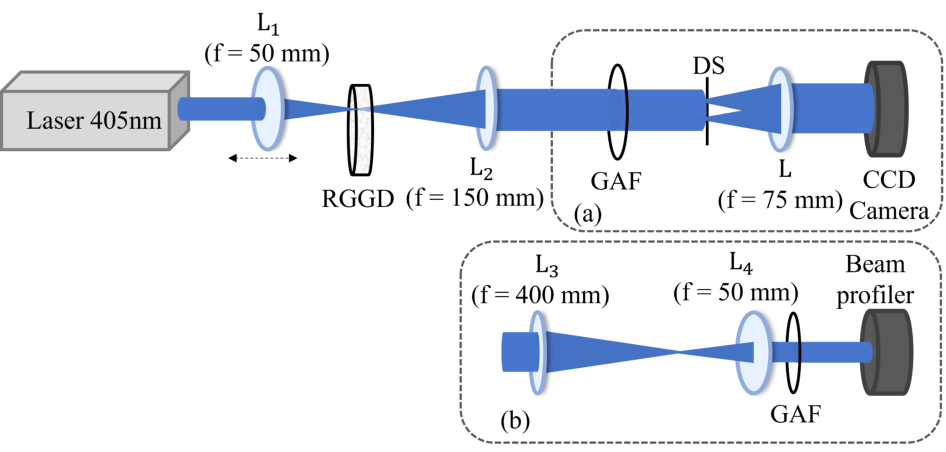}
    \caption{Experimental setup for the generation of Gaussian Schell model (GSM) beam including schemes (a) to measure transverse correlation length and (b) to measure the beam size. Notations: $L$: lens; RGGD: rotating ground glass diffuser; GAF: Gaussian amplitude filter; DS: double-slit, the dashed arrow line represents the translation of $L_{1}$ .}
    \label{pumpsetup}
\end{figure}
\begin{figure}[b!]
    \centering
    \includegraphics[width=\columnwidth]{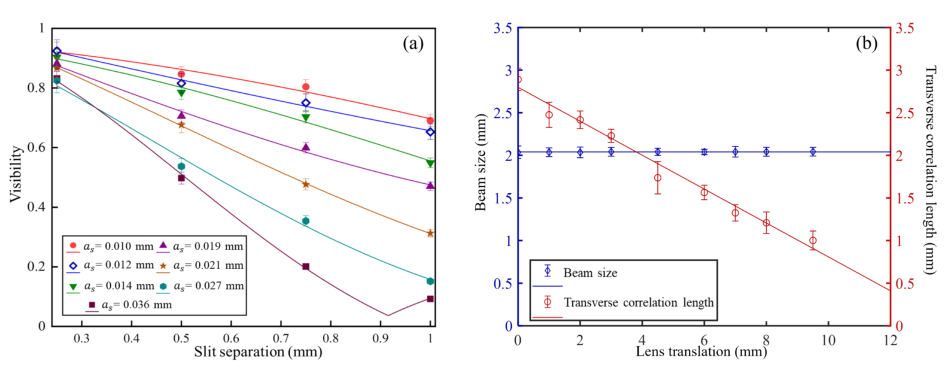}
    \caption{(a) Plot of visibility of interference pattern of GSM pump beam with respect to the slit separation for  variation in spot size at the diffuser. (b) Variation of GSM beam size and transverse correlation length with the translation of lens $L_{1}$.
    Data points correspond to experimental data and solid lines correspond to the fitted theoretical curve.}
    \label{vispump}
\end{figure}
To generate partially spatially coherent qubits, quantum state generation along with a mechanism to control the spatial coherence is required. Our choice for quantum state generation is the SPDC process, in which the coherence properties of the classical pump beam can be controlled. To simplify the analysis, a model for partially coherent beam namely the Gaussian Schell Model (GSM) beam is utilized which offers an analytical formalism to asses the coherence control \cite{Gori20081016,mandel1995optical}. 
The experimental setup shown in figure \ref{pumpsetup} is used for this purpose. A linearly polarized laser beam (405 nm wavelength) is focused on a rotating ground glass diffuser (RGGD) using a lens $L_{1}$ with a focal length of 50 mm to generate an incoherent beam. The spot size on the diffuser is varied with the translation of lens $L_{1}$. Following the van Cittert-Zernike theorem, the incoherent beam acquires spatial coherence as it propagates \cite{mandel1995optical}. It is important to note that the properties of generated biphotons depend on the size of the pump at the non-linear crystal \cite{lee2016,Jeronimo_Moreno2014,preeti2020}. To investigate the sole effect of coherence, the size of the pump beam must remain invariant with the spatial coherence control mechanism. To achieve this, the RGGD is kept at the back focal plane of a lens $L_{2}$ with a focal length of 150 mm, and only the position of lens $L_{1}$ is varied within few millimeters range. The secondary incoherent source thus formed at RGGD changes size with this lens translation and subsequently the emerging beam is partially spatially coherent with size determined by the lens $L_{2}$. The beam is then transformed into a Gaussian Schell model beam by passing it through a Gaussian amplitude filter (GAF) (Figure \ref{pumpsetup}).

The experimental setup to measure the transverse correlation length of the output beam with the translation of lens $L_{1}$ using a double-slit interference experiment is shown in Figure \ref{pumpsetup}(a). Using this setup, we measured the visibility of fringes for four double-slits having same slit width ($a$ =  0.15 mm) and different slit separations ($d$ = 0.25 mm, 0.5 mm, 0.75 mm and 1 mm) (Figure \ref{vispump}(a)). The fringes are captured using a charged couple device (CCD) camera with exposure time of 1 ms. The visibility of the fringes is determined from the respective intensity patterns by making a horizontal scan. It is apparent from figure \ref{vispump}(a) that the visibility of fringes decreases with the increment in spot size at RGGD. The visibility also decreases with the increase in slit separation. For identical intensity at the input of both slits (of DS), the visibility of the interference pattern is a measure of the absolute value of the degree of coherence $|\gamma ({\bf r}_{1},{\bf r}_{2},\tau )|$ due to any two points ${\bf r}_{1}$ and ${\bf r}_{2}$ and is expressed as \cite{hemant2020, Thompson1984, Pearson2018}
\begin{equation}
 \text{Visibility}=|\gamma ({\bf r}_{1},{\bf r}_{2},\tau )|= \Big|  \frac{2 J_{1}(\nu )}{\nu }\Big|, \label{vispumpexp}
\end{equation} 
where $J_{1}(\nu )$ is first order Bessel function, $\nu = \frac{k_{p} d_{12} a_{s}}{f}$ with $k_{p}=\frac{2 \pi}{\lambda_{p}}$ as wavevector of the beam, $d_{12}$ is distance between two points, $a_{s}$ is beam radius (in meters) at the surface of diffuser and $f$ is focal length of lens $L_{2}$. The equation reveals that by measuring the fringe visibility the degree of spatial coherence can be deduced. Experimental data in figure \ref{vispump}(a) fits reasonably well with the theoretical expression given in equation (\ref{vispumpexp}).
The transverse correlation length $l_{c}$ of the beam can be calculated using the expression:
\begin{equation}
l_{c}=\frac{3.832 f}{k_{p}a_{s}}=\frac{3.70503 \times 10^{-8} }{a_{s}} \text{meters}.\label{lc}
\end{equation}
The size of the beam is then reduced by a factor of eight using lenses $L_{3}$ and $L_{4}$ to achieve the resultant GSM pump of desired beam size ($w_{0}$) compatible with the crystal size. Throughout this paper, the degree of spatial coherence is quantified using a parameter $A$, which is expressed as $A=\frac{\delta}{2w_{0}}$ with $\frac{1}{\delta^{2}} = \frac{1}{l_{c}^{2}}+\frac{1}{4 w_{0}^{2}}$ \cite{preeti2021}. The value of $A$ varies from 0 (incoherent beam) to 1 (coherent beam). The range of interest for this study is $0<A<1$ which is the partially coherent regime. The pump beam is subsequently incident on the non-linear crystal to generate the down-converted photons. The variation in transverse correlation length is determined using equation (\ref{lc}), and the resulting change in pump beam size on the crystal's surface with the translation of the lens ($L_1$) is analyzed. In the configuration shown in figure \ref{pumpsetup}(b), the beam profiler is used in place of the crystal to measure the beam size of the pump at the crystal. Figure \ref{vispump}(b) illustrates the influence of lens translation on beam size and transverse correlation length, which is in line with the objectives of our experiment. From this figure, one can clearly deduce that the beam size remains invariant with lens translation and a pump beam of fixed beam size and tunable transverse spatial coherence is achieved at the crystal plane.

\section{\label{sec: Theory-3} Double-slit interference in SPDC photons generated using partially spatially coherent pump}
In the previous section, the spatial coherence features of the pump beam in SPDC have been controlled arbitrarily. A natural question here is, whether these coherence properties translate from the classical pump to the quantum biphotons generated in the SPDC process, as it involves non-linear interactions. In the following, we investigate the coherence properties of biphotons using first order interference, which provides an answer to this question.
\subsection{Theory}
\begin{figure}[h!]
    \centering
    \includegraphics[width=.8\columnwidth]{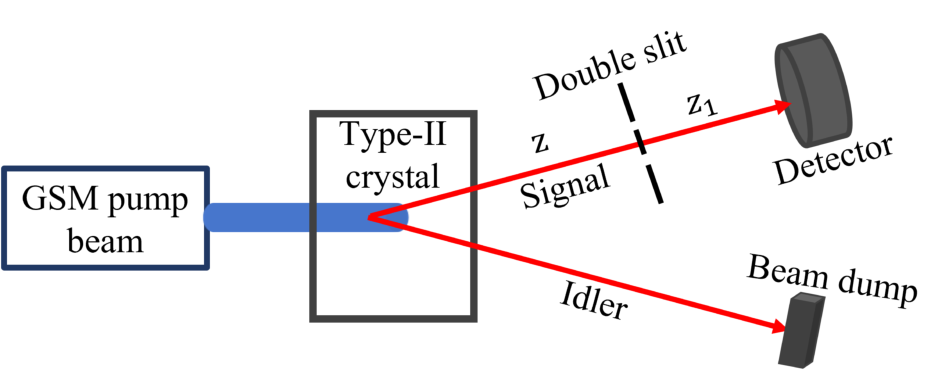}
    \caption{Schematic to measure interference visibility using double-slit in signal beam generated from GSM pump in Type-II SPDC.}
    \label{schematic}
\end{figure}

The theoretical treatment of interference using double-slit in photons generated by type-II non-collinear degenerate SPDC is carried out with the arrangement shown in figure \ref{schematic}. The output state of biphotons after passing through the double-slit, to be detected at positions ${\bf r}_{s}$ and ${\bf r}_{i}$, can be expressed as \cite{Abouraddy2002}
\begin{equation}  
\psi ({\bf r}_{s},{\bf r}_{i}) = \int d{\bf r} t({\bf r}) h_{1}({\bf r}_{s},{\bf r})h_{3}({\bf r}_{i},{\bf r}), \label{state}
\end{equation}
with $h_{1}({\bf r}_{s},{\bf r})$ is impulse response function and $h_{3}({\bf r}_{i},{\bf r})$ contains contribution from phase matching function, pump envelope function and impulse response function. The term $t({\bf r})$ is the transfer function of the double-slit with slit width as $a$ and slit separation as $d$ placed in the path of the generated beam. These terms are expressed as \cite{Saleh2000}
\begin{equation}
\begin{split}
    & h_{1}({\bf r}_{s},{\bf r})=-\frac{\iota}{\lambda_{s}z_{1}}\exp \Big( -\frac{\iota k_{s}}{2 z_{1}} ({\bf r}_{s}-{\bf r})^{2} \Big), \\
     & h_{3}({\bf r}_{i},{\bf r})=\int \int d{\bf q}_{s}d{\bf q}_{i} E_{p}({\bf q}_{s}+{\bf q}_{i})\Phi({\bf q}_{s},{\bf q}_{i}) H_{2}({\bf r},{\bf q}_{s}), \\
     &  H_{2}({\bf r},{\bf q}_{s})=-\frac{1}{2 \pi}\exp \Big( \frac{\iota k_{s}}{ z} (x^{2}+y^{2}) \Big) \exp\Big(-\iota 2 \pi (q_{sx}x  +q_{sy}y) \Big) \exp\Big(\iota \pi \lambda_{s} z (q_{sx}^{2}+q_{sy}^{2}) \Big) ,
\end{split}
\end{equation}
 where $\lambda_{s}$ is wavelength of signal with its wavevector defined by $k_{s}$ and $\iota$ (iota) represents the imaginary part. The signal and idler transverse wavevectors are represented by ${\bf q}_{s}=q_{sx} \hat{x}+q_{sy} \hat{y}$ and ${\bf q}_{i}=q_{ix} \hat{x}+q_{iy} \hat{y}$, respectively, while $z_{1}$ represents the distance of the double-slit from the detector and $z$ represents the distance of the double-slit from the crystal. The electric field of the pump and the phase matching function are denoted by $E_{p}({\bf q}_{s}+{\bf q}_{i})$ and $\Phi({\bf q}_{s},{\bf q}_{i})$ respectively. 
  The interference visibility is determined using the single-photon probability density. It is the probability of detecting signal/idler without detecting the conjugate photon of the generated pair. The single-photon probability density can be expressed as \cite{Ribeiro1999}
\begin{equation}
\begin{split}
    p_{1}({\bf r}_{s})=& \int \int d{\bf r}d{\bf r}^{'} h_{1}({\bf r}_{s},{\bf r}) h_{1}^{*}({\bf r}_{s},{\bf r}^{'})t({\bf r})t^{*}({\bf r}^{'})  h_{3}({\bf r}_{i},{\bf r})h_{3}^{*}({\bf r}_{i},{\bf r}^{'}) .\label{p1_initial}
    \end{split}
\end{equation}
We have considered the GSM beam of size $w_{0}$ and transverse correlation length $l_{c}$ with the expression for pump envelope function expressed in momentum basis as \cite{preeti2021}
 \begin{equation}
\langle E_{p}({\bf q})E^{*}_{p}({\bf q}^{'})\rangle =A_{c}\exp \Big(-b_{1}|{\bf q}|^{2}-b_{1}|{\bf q}^{'}|^{2}+2b_{2}{\bf q}.{\bf q}^{'} \Big) \label{GSM field},
\end{equation}
where $b_{2}=\frac{w_{0}^{2}}{2b_{0}}$, $b_{1}=\frac{(l_{c}+2w_{0})^{2}}{4b_{0}}$, $b_{0}=1+\Big(\frac{l_{c}}{2w_{0}}\Big)^{2}$, $A_{c}=\Big(\frac{A_{p}w_{0}}{2\pi}\Big)^{2}$ and $A_{p}$ is proportionality constant. The phase matching function $\Phi({\bf q}_{s},{\bf q}_{i})$ is $\sinc(\frac{\Delta {\bf q}L}{2})$ and the Gaussian approximation of this function leads to \cite{chan2007}
\begin{equation}
\Phi({\bf q}_{s},{\bf q}_{i})=\sinc\Big(\frac{\Delta {\bf q}L}{2}\Big)\approx \exp\Big(\frac{-\alpha L|{\bf q}_{s}-{\bf q}_{i}|^{2}}{4k_{p}}\Big),\label{phasematching}
\end{equation}
with $\Delta {\bf q}=\frac{|{\bf q}_{s}-{\bf q}_{i}|^{2}}{2k_{p}}$, $L$ be the crystal length and $\alpha=0.455$. Substituting all these expressions in equation (\ref{p1_initial}) yields
\begin{equation}
    \begin{split}
        p_{1}({\bf r}_{s})=&\frac{\iota A_{c}}{2\pi\lambda_{s}z_{1}}\int \int d{\bf r} d{\bf r}^{'} t({\bf r})t^{*}({\bf r}^{'}) \exp \Big[ -\iota k_{s}\Big(\frac{({\bf r}_{s}-{\bf r})^{2}}{2 z_{1}}-\frac{\iota k_{s}}{ z} |{\bf r}|^{2} \Big)\Big]  \int \int d{\bf q}_{s}d{\bf q}_{i} \exp \Big(2\\
        & \times b_{2}{\bf q}.{\bf q}^{'}-b_{1}|{\bf q}|^{2}
        -b_{1}|{\bf q}^{'}|^{2}-\frac{\alpha L|{\bf q}_{s}-{\bf q}_{i}|^{2}}{4k_{p}} \Big)  \exp\Big(\iota \pi ( \lambda_{s} z |{\bf q}|^{2}- 2  {\bf q}_{s}.{\bf r}) \Big) \label{p1}
    \end{split}
\end{equation}
For simplicity, the expression is reduced to 1-D (x-axis). The transfer function of the double-slit is given by
\cite{joshi2020spatial}
\begin{equation}
   t(x)= \begin{cases}
      1 & \text{ $\frac{d-a}{2} \leq x \leq \frac{d+a}{2}$ and $-\big(\frac{d+a}{2}\big) \leq x \leq -\big(\frac{d-a}{2}\big)$}\\
      0 & \text{otherwise}.
    \end{cases} 
  \end{equation}
Substituting the value of the transfer function for a double-slit, the final expression for single photon probability density is obtained. The final expression is numerically integrated to get the intensity profile at the detector position. The intensity depends on the slit width, slit separation, the degree of spatial coherence of the pump beam, and the distance of the double-slit from the crystal and the detector, respectively.
\subsection{Experiment}
\begin{figure}[ht!]
    \centering
    \includegraphics[width=.9\columnwidth]{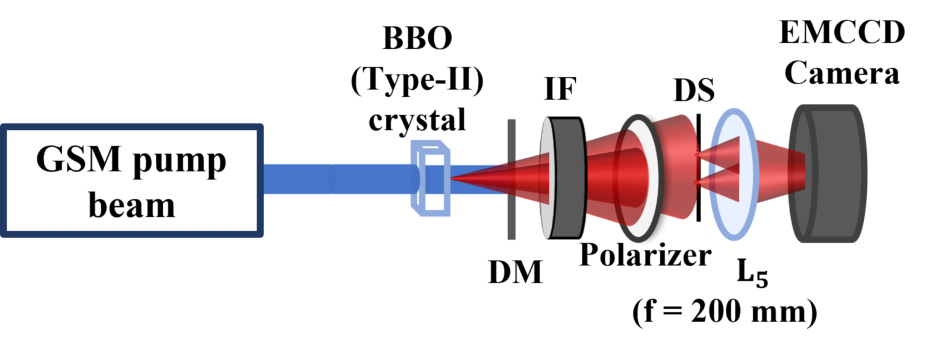}
    \caption{Experimental setup for the double-slit interference in photons generated using non-collinear degenerate type-II SPDC with GSM pump beam. Notations: $L_{5}$: lens; DM: dichroic mirror; IF: interference filter; DS: double-slit.}
    \label{fig8}
\end{figure}
\begin{figure}[t!]
    \centering
    \includegraphics[width=.9\columnwidth]{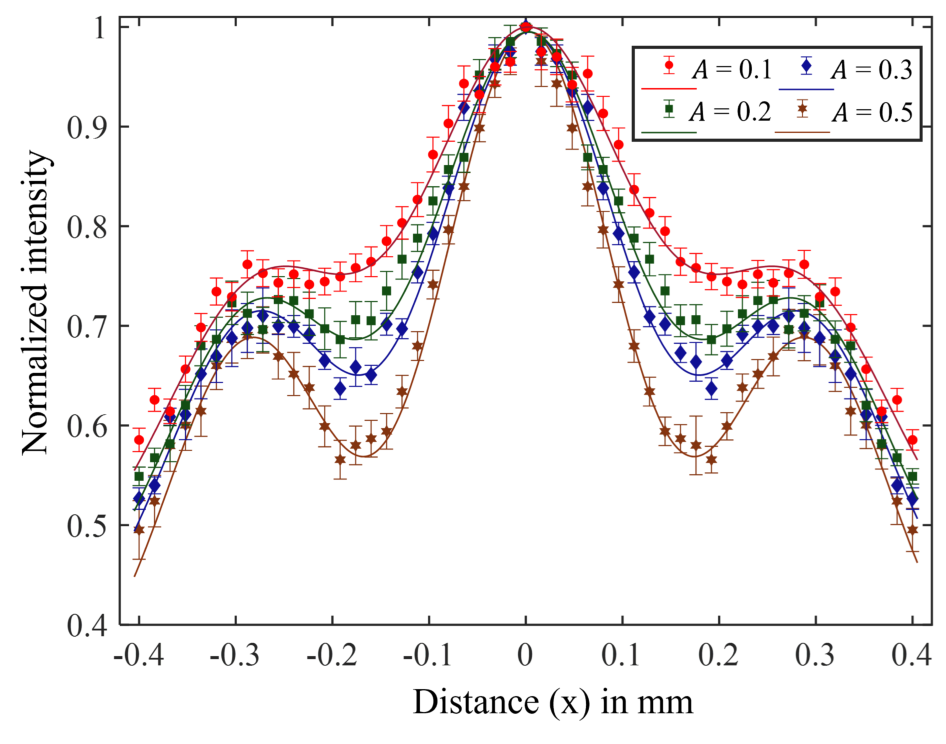}
    \caption{Intensity profile of interference fringes obtained for the signal beam captured using EMCCD camera with variation in spatial coherence of pump ($A$). Markers with error bars correspond to experimental data and solid lines are fitted theoretical curves. } 
    \label{intensitysignal}
\end{figure}

 The experimental setup to investigate the double-slit interference of photons produced in the non-collinear degenerate type-II SPDC process is depicted in figure \ref{fig8}. 
 The generated partially spatially coherent pump beam (see Section \ref{sec:Theory-1}) is incident on type-II beta Barium Borate (BBO) crystal which generates down-converted photons represented by two cones of orthogonal polarizations. The residual pump beam is blocked using a dichroic mirror (DM). The interference filter (IF) with a central wavelength at 810 nm and bandwidth of 10 nm is used for the  spectral selection of biphotons. The selection of the vertically polarized signal beam for the double-slit interference is done by filtering the idler beam from the output of type-II SPDC.  This is done with the aid of a polarizer  
 placed at $90^{\circ}$ with respect to the pass axis which separates the ordinary polarized beam (signal) from the extraordinary beam (idler). The ordinary beam (signal) is then directed through a double-slit of 0.15 mm slit width ($a$) placed before a 200 mm focal length lens ($L_{5}$). Using an electron-multiplying charge-coupled device (EMCCD) set in a photon counting setting, 20000 images are accumulated with an exposure time of 20 ms to record the interference pattern of the output beam. A zero point or reference point is designated to a pixel position of the signal in the $x$ direction. Image processing is used to extract the intensity of each individual pixel in an image.
 \begin{figure}[t!]
    \centering
    \includegraphics[width=.9\columnwidth]{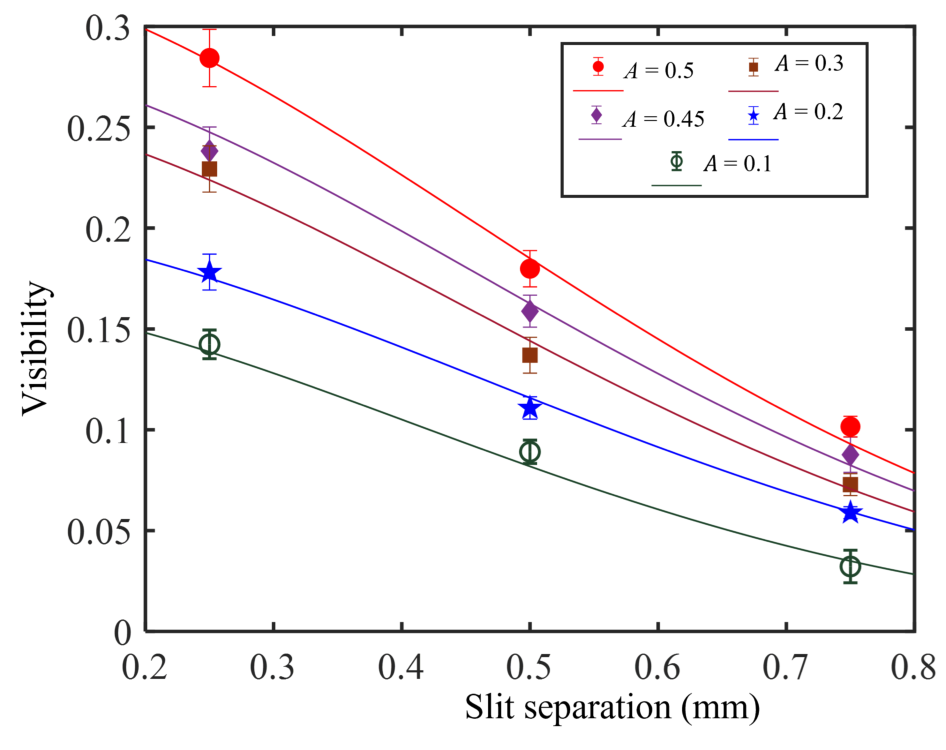}
    \caption{Plot of visibility of interference patterns of photons vs slit separation (of double-slit) for different values of spatial coherence of pump. Data points represent  experimental observation and solid lines correspond to the fitted theoretical curve. }
    \label{visibility signal}
\end{figure}

The experimentally observed intensity at interference plane is normalized and fitted with the theory as shown in figure \ref{intensitysignal}. This figure shows the interference pattern for various values of pump spatial coherence with double-slit having 0.25 mm slit separation ($d$). The achieved interference patterns degrade with the decrease in the spatial coherence of the pump (A). As indicated by the solid curves, the theoretically expected patterns using equation (\ref{p1}) are found in good agreement with the observed data points (markers with error bars). The resulting visibilities of the interference patterns for different values of spatial coherence of the pump are obtained for changing the slit separation (0.25 mm, 0.5 mm, and 0.75 mm) with the same slit width (0.15 mm) and are plotted in figure \ref{visibility signal}. The experimentally observed values in the figure are fitted with a Gaussian function. It is evident that when the slit separation increases, the visibility of the interference pattern diminishes, which is consistent with the theoretical expression within the experimental errors. Furthermore, the visibility decreases with a decrease in the spatial coherence of the pump. The results show that the correlations in photons generated using the GSM pump beam in the SPDC process are directly proportional to the spatial coherence of the pump beam. This is the key result of the paper as it shows that the spatial coherence is transferred from the pump (classical) to biphotons (quantum field). Thus, the partially spatially coherent qubits with a tunable degree of spatial coherence are generated which could be used in free space optical communication due to robustness against atmospheric turbulence.  

\section{\label{sec:Theory-2} Transverse distribution of twin photons generated in non-collinear configuration}
In addition to the effect of the spatial coherence of the GSM pump on double-slit interference, the distribution of generated photons in the transverse plane is observed for type-I and type-II non-collinear configurations. The spatial profiles are measured using the scheme shown in figure \ref{fig8}, with the exception of the polarizer and double-slit.

\begin{figure*}[ht!]
    \centering
\includegraphics[width=.94\linewidth,height=.6\linewidth,keepaspectratio]{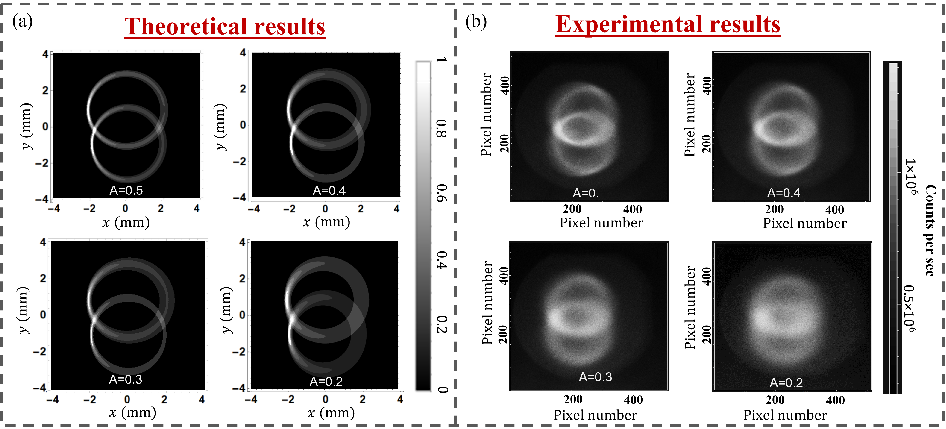}
   \caption{(a) Simulated and (b) experimentally observed spatial distribution of output beam generated in type-II SPDC for different values of spatial coherence of pump ($A$).}
   \label{type2}
\end{figure*}
\subsection{Type-II SPDC}
The biphotons are generated in the type-II SPDC process with a GSM pump using beta Barium Borate (BBO) crystal in a non-collinear configuration. The BBO crystal of 2 mm crystal length interacts with a horizontally polarized pump with a wavelength of 405 nm, which produces frequency degenerate biphotons in type-II non-collinear geometry. The spectral selection of the biphotons using an interference filter (IF) with a bandwidth of 10 nm and a central wavelength of 810 nm is the same as is discussed in the former section. Utilizing a lens $L_5$ with a 200 mm focal length, the output is collimated before being detected by an EMCCD camera (Figure \ref{fig8}).

\begin{figure}[hb!]
\centering
\includegraphics[width=\columnwidth,height=\linewidth,keepaspectratio]{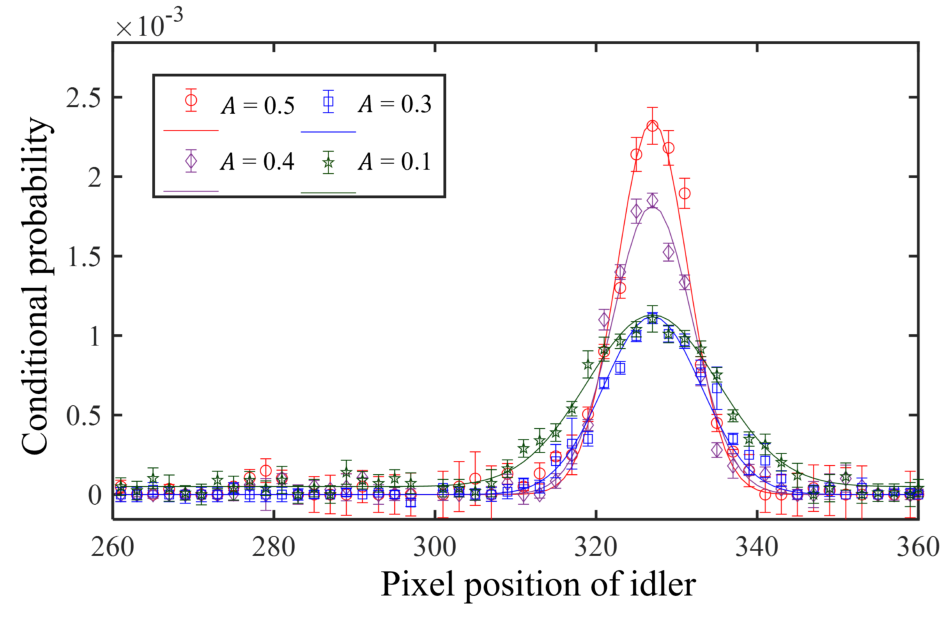}
   \caption{Conditional detection probability of idler in the horizontal direction with variation in spatial coherence of pump. The vertical pixel position for both signal and idler beams, and the horizontal pixel position of the signal are fixed. The markers represent experimental values and are fitted with theoretical curves (solid lines). }
    \label{coinc}
\end{figure}
Ordinary signal (V polarized) and extraordinary idler (H polarized) photons are created by the extraordinary pump (H polarized). The extraordinary pump and idler encounter spatial walk-off in this configuration. 
By using a photon counting setting with an exposure time of 10 ms and integrating 20,000 images, the experimental profiles are acquired (figure \ref{type2}(b)). The corresponding simulated profiles are shown in figure \ref{type2}(a). In the profiles, the two rings indicate ordinary and extraordinary polarizations. The location of the polarization-entangled photons is at the points where these two rings overlap. With the degradation of the spatial coherence of the pump, the rings broaden asymmetrically due to the spatial walk-off of the pump beam and idler beam and the overlap region becomes fuzzier. This is consistent with the theoretical formulation within experimental limits \cite{preeti2020}.

The conditional probability of idler \cite{Hugo2018} with signal fixed at a particular position is also calculated and is shown in figure \ref{coinc}. The coincidence counts between $i$ and $j$ pixel position in captured images is calculated using $C=\langle n_{i}n_{j} \rangle -\langle n_{i}\rangle \langle n_{j}\rangle $ \cite{Avella2016} where $n_{i}$ and $n_{j}$ are average counts at position $i$ and $j$ respectively. The signal is fixed at pixel position $x_{s},y_{s}$ 
and $y$ position of idler ($y_{i}$) is also fixed
. These positions are the two points at which polarization entanglement can be observed. The scan of coincidence counts of idler along $x$ axis with signal conditioned at one of the entangled positions is fitted with Gaussian distribution. The increase in full-width half maxima (FWHM) of the fitted curve indicates the increase in the area of correlation at the observational point. It is clear from figure \ref{coinc} that with a decrease in spatial coherence of the pump, the strength (conditional probability) of correlations decreases and the area of spatial correlation broadens (increase in FWHM). This broadened distribution infers correlation with more number of spatial modes (multi-mode nature) as the partial coherence of the signal beam increases. Owing to this multi-mode nature, alike classical partially coherent beams, these partially coherent qubits are expected to show robustness against atmospheric turbulence and higher detection probability on free-space propagation \cite{gbur2014partially}. Similarly, with a rise in coherence of pump, the conditional probability increases and the distribution becomes narrower, resulting in stronger correlations with fewer spatial modes (single mode nature) of coherent beams. This is another significant outcome of this study.

\subsection{Type-I SPDC}
\begin{figure*}[hb!]%
\centering
\includegraphics[width=.94\linewidth,height=.6\linewidth,keepaspectratio]{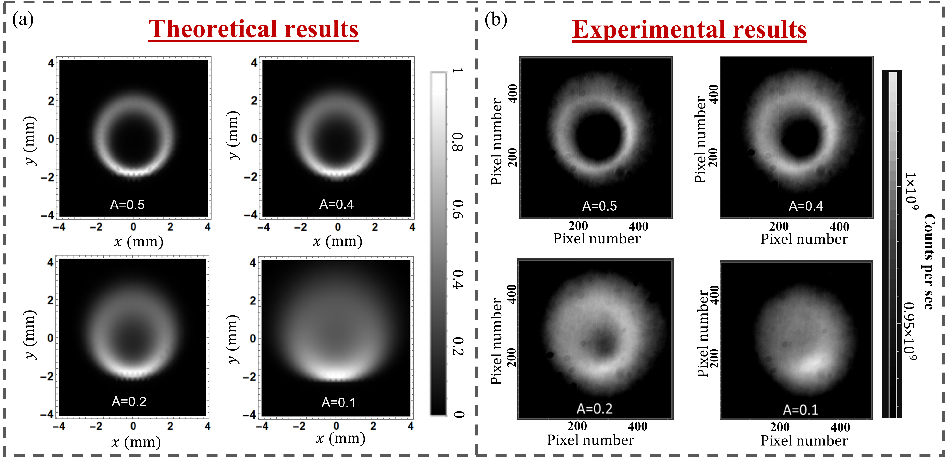}
   \caption{Spatial profiles of biphotons generated in type-I non-collinear SPDC for different values of spatial coherence of pump ($A$). (a) Simulated results (b) Experimental observations.}
    \label{type1}
\end{figure*}

In this case, Bismuth Borate (BiBO) crystal is employed as non-linear medium to generate the biphotons in the type-I degenerate SPDC process in non-collinear geometry.
The spatial profile of these biphotons is captured using an EMCCD camera using the experimental setup depicted in figure \ref{fig8}, excluding the polarizer and double-slit, and with bi-axial BiBO (type-I) crystal replacing BBO (type-II) crystal.
The extraordinary pump (H polarized) generates ordinary polarized (V polarized) photons. The simulated and experimental results of spatial profile in non-collinear type-I SPDC with variation in the degree of spatial coherence of pump are shown in figure \ref{type1}(a) and (b) respectively. 
The ring displays the biphoton distribution along the pump's transverse axis. The profiles are recorded for various values of the pump's spatial coherence with lens translation $L_{1}$. A careful look into the profiles reveals the broadening of SPDC rings in a particular direction, along with a decrease in the transverse correlation length of the pump beam. This verifies the theoretical framework proposed in our earlier study that spatial correlations in twin photons deteriorate with the loss in spatial coherence of the pump \cite{preeti2020}, and shows asymmetric broadening in the opposite direction of spatial walk-off of the pump. These results show that one can generate a robust beam with variable spatial coherence, which could be applicable for communication purposes. The study also reveals that the partial spatial coherence should be optimized  carefully so that it does not affect other features such as polarization entanglement to a great extent.
\section{\label{sec:Conclusion}Conclusion}
The article illustrates an experimental method to generate the spatial coherence controlled qubits using type-I and type-II non-collinear SPDC processes. The classical double-slit interference method is used to characterize the spatial coherence of photons generated in type-II non-collinear SPDC pumped with a GSM beam having a tunable degree of spatial coherence. The experimental results are verified with the theoretical formulation for different values of slit separation and variation in spatial coherence of the pump for a fixed beam size. This study shows the direct dependence of the spatial coherence of the pump on the visibility in the interference pattern of biphotons. In addition, the spatial profiles of down-converted photons in type-I and type-II SPDC in non-collinear geometry are experimentally observed with a partially spatially coherent pump beam. The asymmetric broadening of SPDC rings in the direction opposite to the spatial walk-off of the extraordinary beam clearly depicts the significant impact of the spatial coherence of the pump on generated photons. The study of the conditional probability of spatial correlations at the location of polarization entanglement shows the enhancement in spatial modes with degradation in pump coherence. 
This implies that the strong correlation at two conjugate points expected with a coherent pump decreases with a decrease in the spatial coherence of the pump. The polarization entanglement would also be impacted by the decreased spatial correlations and multiple spatial modes in type-II SPDC. These findings are crucial in selection of appropriate parameters of a partially coherent pump beam that will produce entangled photon sources while maintaining a high degree of spatial correlations. 
The robust nature of these partially spatially coherent qubits against atmospheric turbulence makes them useful for long-distance free-space communication, quantum teleportation, and reduction in speckles for quantum imaging.

\ack
The authors acknowledge the funding received from the Department of Science and Technology (DST), India under the QUEST scheme (DST/ICPS/QUEST/Theme-I/2019) to carry out this work. Author PS and SR acknowledge financial support from the Council of Scientific and Industrial Research (CSIR) India and DST INSPIRE, respectively for Senior Research Fellowship (SRF). The authors are grateful to Hemant Kumar Singh for his suggestions in the experimental section involving pump generation. \\
\newline
{\bf Conflicts of interest}
The authors declare that they have no conflict of interest.\\
\newline
{\bf Data Availability Statement} The data that support the findings of the study are available upon request from the authors.
\section*{References}

\bibliographystyle{iopart-num}

\end{document}